\documentclass[aps,amssymb,amsmath,tightenlines]{revtex4}
\usepackage{graphicx}
\newcommand{\ket}{\rangle}
\newcommand{\bra}{\langle}

\begin{document}

\pagestyle{empty}

\title{An Extension of the Wigner-Araki-Yanase Theorem to Multiplicative
Conserved Quantities }

\author{Bernhard K. Meister}

\affiliation{Physics Department, Renmin University of China, Beijing
100872,\\
 China 
   }

\begin{abstract}
An extension of the Wigner-Araki-Yanase theorem to multiplicative
conserved quantities is presented and approximate versions of the
theorem are discussed. The theorem proposed  by Wigner, and
subsequently proven by Araki and Yanase in a general setting, deals
with the impossibility of exact nondestructive measurements of
observables that do not commute with additive conserved quantities.
An analogous theorem concerning the limitation of nondestructive
measurements of multiplicative conserved quantities is proven in
this paper. The result is analyzed in the context of earlier work.
An approximate form of the theorem, more appropriate in experimental
settings, as well as possible extensions are also briefly discussed.
\end{abstract}
\maketitle
 \setcounter{section}{0}

\section{Introduction}

An extension of the Wigner-Araki-Yanase theorem is presented and an
approximate version of the theorem is discussed. The theorem
proposed by Wigner~\cite{wig} following an earlier paper by
L\"uders\cite{luders}, and subsequently proven by Araki \&
Yanase~\cite{araki}, deals with the impossibility of exact
nondestructive quantum mechanical measurements of observables that
do not commute with additive conserved quantities. An analogous
theorem concerning the limitation of nondestructive measurements of
multiplicative conserved quantities that are related to discrete
symmetries is proven here. The result is analyzed in the context of
earlier work. An approximate form of the theorem, more appropriate
in experimental settings, is also briefly analyzed.

The intense study of quantum computers  has led to a revival of
interest in foundational issues in quantum mechanics including
various aspects of the measurement process.
One result in this context is the further
analysis of  approximate versions of the Winger-Arkai-Yanase
theorem. This work has been carried out in particular by
Ozawa~\cite{ozawa1, ozawa2}, who has discussed implications of the
theorem for the construction of quantum gates.

Let us restate the Wigner-Araki-Yanase theorem briefly. An additive
observable  acting on the combined Hilbert space $H_1 \otimes H_2$,
where $H_1$ and $H_2$ are the Hilbert spaces associated with the
measured object and the measurement apparatus respectively,   has
the following form $\hat{L}^A\otimes \hat{1} + \hat{1}\otimes
\hat{L}^B$, where $\hat{L}^A$ is an observable\footnote{All the
observables in this paper are assumed to have a finite discrete
non-degenerate spectrum.} of the observed system and $\hat{L}^B$
describes the measuring apparatus.  This additive observables is a
{\em conserved} quantity, which is generally associated with a
continuous symmetry, if it satisfies the following equation
\begin{eqnarray}
\hat{L}^A\otimes \hat{1} + \hat{1}\otimes \hat{L}^B=
\hat{U}^{\dagger}(\hat{L}^A\otimes \hat{1} + \hat{1} \otimes
\hat{L}^B)\hat{U}, \label{eq:1}
\end{eqnarray}
 where $\hat{U}$ is the evolution operator of the combined system. The previous
 equation (\ref{eq:1}) is equivalent to  $[\hat{U},\hat{L}^A\otimes \hat{1} + \hat{1}\otimes
 \hat{L}^B]=0$.
 The Wigner-Araki-Yanase theorem now states  under some reasonable assumptions about the observable and the Hilbert spaces
 that one can only measure an observable
 $\hat{O}\otimes \hat{1}$ of the system in a nondestructive way, if
 $[\hat{O},\hat{L}^A]=0$. For clarification, a
 nondestructive measurement is defined to be a measurement, which
 leaves the system in its original eigenstate. As a further clarification, this does not violate the non-cloning
 theorem, since the measurement is only nondestructive for the
 preselected basis of the original Hilbert space $H_1$ and not for any superposition of basis elements.

In the following section the Wigner-Araki-Yanase theorem is extended
in a novel way to multiplicative conserved quantities.
A description of an approximate version of the theorem is presented
in a subsequent section, which leads to an inequality similar in
nature, but different in form, to the one already established by
Ozawa. The paper is rounded off by a brief speculation about further
extensions of the theorem.

\section{The Wigner-Araki-Yanase theorem for multiplicative conserved quantities}
\label{sec:6}

In the following paragraphs a theorem for multiplicative conserved
quantities is proven that extends the Wigner-Araki-Yanase theorem.
The notation introduced above is employed without modification
wherever possible. A multiplicative quantity can be written as
$\hat{L}^A\otimes \hat{L}^B$. It is a multiplicative {\em conserved}
quantity, if
\begin{eqnarray}
\hat{L}^A\otimes \hat{L}^B= \hat{U}^{\dagger}(\hat{L}^A\otimes
\hat{L}^B)\hat{U}. \label{eq:201}
\end{eqnarray}
Now we can proof under some reasonable assumptions - i.e both
Hilbert spaces are finite dimensional with $H_1$ $n_1$-dimensional
and $H_2$ less than $2n_1$-dimensional, the rank of $\hat{L}^B$ is
maximal, and all the eigenvalues of $\hat{L}^A$ and $\hat{L}^B$
are positive- that every observable $\hat{O}$ that does not
satisfies $[\hat{O},\hat{L}^A]=0$ cannot be measured exactly in a
nondestructive way.
Let us next state the theorem precisely and then give the proof.
\vspace{.41cm}

 {\em Theorem:}
If an operator $\hat{O}$ can be measured exactly in a
nondestructive way, then it has to satisfy  the property
$[\hat{O},\hat{L}^A]=0$ under the assumption that
$\hat{L}^A\otimes \hat{L}^B$ is a multiplicative conserved
quantity, i.e. $U^{\dagger} (\hat{L}^A\otimes \hat{L}^B)
U=\hat{L}^A\otimes \hat{L}^B$, $\hat{L}^B$ has maximal rank, the
Hilbert spaces of the system $H_1$ is $n_1$-dimensional, and the
Hilbert space of the measurement apparatus $H_2$ is less than
$2n_1$-dimensional. We further assume that all eigenvalues of
$\hat{L}^A$ and $\hat{L}^B$ are greater than zero.

\vspace{.31cm}

{\em Proof:} Consider $\bra u(i) | \hat{L}^A| u(j)\ket\bra v |
\hat{L}^B| v\ket$  for any $i$ $\&$ $j$ between $1$ and $n_1$, which
transforms to
\begin{eqnarray}
&&\bra u(i)\otimes v|\hat{L}^A\otimes \hat{L}^B|u(j) \otimes v\ket \\
&=& \bra u(i)\otimes v|\hat{U}^{\dagger}(\hat{L}^A\otimes
\hat{L}^B)\hat{U}|
u(j)\otimes v\ket \\
& =& \bra u(i)|\hat{L}^A|u(j) \ket \bra v(i)|\hat{L}^B |v(j)\ket,
\label{eq:34}
\end{eqnarray}
since $\hat{L}^A\otimes \hat{L}^B$ is a conserved quantity. The
equality
\begin{eqnarray}
\bra u(i) | \hat{L}^A| u(j)\ket\bra v | \hat{L}^B| v\ket = \bra
u(i)|\hat{L}^A|u(j) \ket \bra v(i)|\hat{L}^B |v(j)\ket
\end{eqnarray}
 can
either be satisfied, if $\bra v(i) | \hat{L}^A| v(j)\ket=$ const or
by $\bra u(i) | \hat{L}^A| u(j)\ket=0$. The second case
immediately implies the theorem.
Therefore, we only have to show that the first case leads to a
contradiction. This can be shown in the following way. First we
expand $\hat{L}^B|v(j)\ket$ in terms of the basis for $|v\ket$:
\begin{eqnarray}
\hat{L}^B|v(j)\ket = \sum_{k=1}^{n_2} L_{kj}^B |v(k)\ket
\label{eq:35}
\end{eqnarray}
Next we exploit the orthogonality of the basis for $|v\ket$ to get:
$\bra v(i) | \hat{L}^B| v(j)\ket=L^B_{ij}$. From this it follows
that the sub-matrix of   $L^B_{ij}$ with $i,j\in{1,...,n_1}$  is at
most of rank one. The full matrix $L_{ij}^B$ is of maximal rank, as
stated in the assumptions, implying   that the sub-matrix $L_{ij}^B$
with $i\in{1,...,n_1}$ and $j\in{n_1+1,...,n_2}$ is
$n_1-1$-dimensional. This is only possible, if  $n_2$     is at
least $2n_1$, which contradicts one of the assumptions. Therefore,
$\bra u(i) | \hat{L}^A| u(j)\ket=0$ for all $i,j\in{1,...,n_1}$, and
like in the proof of the original Wigner-Araki-Yanase this implies
$[\hat{O},\hat{L}^A]$. ${\bf QED}.$

\vspace{.1cm}

 Examples for
multiplicative conserved quantities can be found in particle
physics, where parity is a multiplicative conserved quantity, and
other areas where discrete symmetries are relevant.

\section{An Approximate Version of the Inequality}

In this section we provide an approximate version of the theorem
proven above in terms of an inequality. For practical purposes, i.e.
to be able to point out theoretical limits in experiments, it is
necessary to go beyond a no-go theorem and deal with approximate
versions of the statement. Particular interest in areas like quantum
information theory has been shown for measurements which are
nondestructive but only distinguish the potential input states with
less than $100\%$ accuracy by correlating the measurement apparatus
imperfectly with the state of the system. One natural way, pioneered
by Ozawa \cite{ozawa1}, is to introduce a {\it Noise operator}
$\hat{N}$ defined as
\begin{eqnarray}
\hat{N}:=\hat{M}(t+\Delta t)-\hat{O}(t),  \label{eq:35}
\end{eqnarray}
where $\hat{M}(t+\Delta t)$ is the potentially error-prone probe
observable, i.e. $\bra v(i)|v(j) \ket$ is not necessarily zero.
One could, of course, pursue other approaches, e.g. for example
assume that either the measurement is not exactly nondestructive or
that the conserved quantity is not perfectly conserved. These other
approaches will be discussed in a separate paper\cite{meister}. Here
we follow closely  the reasoning of  Ozawa, who derived in the
additive case an inequality for a quantity, $\epsilon (\psi)=\bra
N^2 \ket$ associated with the state-dependent measurement noise,
related to the noise operator.
After some simple
manipulations we can show that the measurement noise
$\epsilon(\psi)$ is bounded below by a function of the various other
operators involved. The inequality has the form
\begin{eqnarray}
\epsilon (\psi)^2\geq \frac{|\bra [\hat{O}(t),\hat{L}^A]\otimes
\hat{L}^B - \hat{L}^A\otimes [\hat{M}(t+\Delta t),\hat{L}^B]\ket|
^2}{4(\Delta \hat{L}^A)^2 (\Delta \hat{L}^B)^2} .
\end{eqnarray}

In the derivation of the inequality we utilized the Cauchy-Schwartz
inequality, $\epsilon (\psi)\geq (\Delta N)^2 $, and  that $\Delta [
\hat{L}^A\otimes \hat{L}^B]^2=(\Delta  \hat{L}^A \Delta
\hat{L}^B)^2$. The last property follows from the fact that the
variance of product states is the product of the variances of its
components.

To simplify the inequality we assume the Yanase
condition\cite{yanase}: $[\hat{M}, \hat{L}^B]=0$. The inequality
then takes the form
\begin{eqnarray}
\epsilon (\psi)^2\geq \frac{|\bra
[\hat{O}(t),\hat{L}^A]\otimes\hat{L}^B \ket |^2}{4(\Delta
\hat{L}^A)^2 (\Delta \hat{L}^B)^2 }.
\end{eqnarray}
If the expectation value of $\hat{L}^B$ is zero this further
simplifies to  \begin{eqnarray} \epsilon (\psi)^2\geq \frac{|\bra
[\hat{O}(t),\hat{L}^A]\ket |^2}{4(\Delta \hat{L}^A)^2 }.
\end{eqnarray}
Unlike in the additive case in this very special multiplicative case
with $\bra \hat{L}^B\ket=0$  a decrease in this particular form of
the lower bound for the measuring noise cannot simply be achieved by
increasing the variance of $\hat{L}^B$.

\section{Extensions}

There are various extensions one ought to  consider. One particular
interesting case is to establish the Wigner-Araki-Yanase Theorem for
generalized conserved quantities with both additive as well as
multiplicative components. This extension is important, when one
considers for example a generalized interaction between system and
environment, and might be testable experimentally.

Also of interest would be to explore the relationship between
multiplicative conserved quantities and the implementation
limitations for quantum computers. Of particular interest is also to
explore in general the relation between additive and multiplicative
version of the Wigner-Araki-Yanase theorem. This and other
extensions can be found in an enlarged version of the article, which
is under preparation \cite{meister}.

\section{Acknowledgements}
The author gratefully acknowledges clarifying discussions with
Prof. M. Ozawa and Dr. G. Kimura from Tohoku University both at
and after the QCMC2006 meeting.
Financial Assistance of the NSF of China is gratefully acknowledged.

\end{document}